\documentclass[10pt]{article}
\usepackage{graphicx}
\usepackage{amsmath}
\usepackage{amssymb}
\usepackage{caption2}
\begin{document}
\begin{center}
{\large {\bf \sc{Three-body strong decays of the $Y(4230)$  via the light-cone QCD sum rules  }}} \\[2mm]
Zhi-Gang  Wang \footnote{E-mail: zgwang@aliyun.com.  }     \\
 Department of Physics, North China Electric Power University, Baoding 071003, P. R. China
\end{center}

\begin{abstract}
We tentatively assign the $Y(4230)$ as the vector tetraquark state with a relative P-wave between the scalar diquark pair, and  explore the three-body strong decays $Y(4230) \to \bar{D}^{*-}D^{*0}\pi^+$, $\bar{D}^{*-}D^0\pi^+$,  $J/\psi\pi^+\pi^- $ and  $ J/\psi K^+K^-$ with the light-cone QCD sum rules by assuming contact four-meson coupling constants. The resulting partial decay widths  are too small to account for the experimental data, and we expect those decays take place through an intermediate meson. We can search for the intermediate states and precisely measure the branching fractions to diagnose the nature of the $Y$ states.
 \end{abstract}



\section{Introduction}
In the past years,  several vector  charmonium-like states have been observed,  they  cannot be accommodated comfortably  in the traditional  charmonia.

In 2005, the BaBar collaboration  investigated  the initial-state radiation (ISR) process  $e^+ e^- \to \gamma_{ISR} \,\pi^+\pi^- J/\psi$ and observed the $Y(4260)$   in the $\pi^+\pi^- J/\psi$ mass spectrum  \cite{BaBar4260-0506}, subsequently,  the $Y(4260)$ was confirmed by the Belle and CLEO collaborations \cite{Belle-0707,CLEO-0606}.

In 2006, the BaBar collaboration observed  a broad structure at  $4.32\,\rm{GeV}$ in the $\pi^+\pi^- \psi^\prime$ mass spectrum  in the process $e^+e^- \to \pi^+ \pi^- \psi^{\prime}$ \cite{BaBar-Y4360}.
In 2007, the  Belle collaboration  studied the process $e^+e^- \to \gamma_{ISR}\pi^+ \pi^- \psi^{\prime}$, and  observed two structures $Y(4360)$ and $Y(4660)$ in the $\pi^+ \pi^- \psi^{\prime}$ mass spectrum \cite{Belle4660-0707-1,Belle4660-0707-2}.
In 2008, the Belle collaboration explored   the  process $e^+e^- \to \gamma_{ISR} \Lambda_c^+ \Lambda_c^-$   and observed the $Y(4630)$   in the $\Lambda_c^+ \Lambda_c^-$  mass spectrum \cite{Belle4630-0807}. The $Y(4360)$ and $Y(4660/4630)$ were confirmed by the BaBar collaboration \cite{BaBar-Y4360-Y4660}.

In 2014, the BESIII collaboration searched for the process  $e^+e^-\to \omega\chi_{c0/1/2}$, and observed a resonance $Y(4220)$ in the $\omega\chi_{c0}$ cross section,  the measured mass and width  are $4230\pm 8\pm 6\, \rm{ MeV}$    and $ 38\pm 12\pm 2\,\rm{MeV}$, respectively \cite{BES-2014-4230}.
In 2016, the BESIII collaboration measured the cross sections of the process $e^+ e^- \to \pi^+\pi^- h_c$, and observed two resonances, the $Y(4220)$ has a mass of $4218.4^{+5.5}_{-4.5}\pm0.9\,\rm{MeV}$ and a width of $66.0^{+12.3}_{-8.3}\pm0.4\,\rm{MeV}$, respectively, and the $Y(4390)$ has a mass  of $4391.6^{+6.3}_{-6.8}\pm1.0\,\rm{MeV}$ and a width of $139.5^{+16.2}_{-20.6}\pm0.6\,\rm{MeV}$, respectively \cite{BES-Y4390}.
Also in 2016, the BESIII collaboration precisely measured the cross section of the process $e^+ e^- \to  \pi^+\pi^- J/\psi$   and observed  two resonances, which are consistent with the  $Y(4230)$ and $Y(4360)$, respectively \cite{BES-Y4220-Y4320}.

In 2018, the BESIII collaboration measured  the cross section of the process $e^+e^-\to \pi^+D^0\bar{D}^{*-}$ and
 observed two enhancements around 4.23 and 4.40 $\rm{GeV}$, respectively, the lower enhancement has a mass of $4228.6 \pm 4.1 \pm 6.3 \,\rm{MeV}$ and a width of $77.0 \pm 6.8 \pm 6.3 \,\rm {MeV}$, and it is compatible with the $Y(4230)$ \cite{BESIII-DDvpi}.
In 2022, the BESIII collaboration explored the $e^+e^-\to K^+K^-J/\psi$ cross sections  and observed two resonant structures, one is consistent with the well-known $Y(4230)$; the other was  observed for the first time and denoted as the $Y(4500)$ \cite{BESIII-KK-4500}.

Recently, the BESIII collaboration explored the Born cross sections of the process  $e^+e^-\to \bar{D}^{*-}D^{*0}\pi^+$ and observed three enhancements, whose  masses are $4209.6\pm 4.7\pm 5.9\,\rm{MeV}$, $4469.1\pm26.2\pm3.6\,\rm{MeV}$ and $4675.3\pm 29.5\pm 3.5\,\rm{MeV}$, respectively,  and widths are $81.6\pm 17.8\pm 9.0\,\rm{MeV}$, $246.3\pm 36.7\pm 9.4\,\rm{MeV}$ and $218.3\pm 72.9\pm 9.3\,\rm{MeV}$, respectively, and they are consistent with the $Y(4230)$, $Y(4500)$ and $Y(4660)$ states, respectively   \cite{X4500-BESIII}.

There have been several assignments for those $Y$ states, such as the tetraquark states \cite{Maiani-4260,Maiani-II-type,Ali-Maiani-Y,Brodsky-PRL,Vector-Tetra-WZG-P-wave-1,
Vector-Tetra-WZG-P-wave,ZhangHuang-PRD,ChenZhu,WZG-Vector-NPB,
Nielsen-4260-4460,WangEPJC-4660,WangEPJC-1601,WangY4360Y4660-1803,Vector-Tetra-Ivanov,Vector-Tetra-Lebed}, hybrid states \cite{Hybrid-4260-1,Hybrid-4260-2,Hybrid-4260-Lattice,BO-potential}, hadro-charmonium states \cite{GuoFK-4660-psif0,WangZG-4660-psif0},
 molecular states \cite{Mole-DingGJ,Zhao-PRL-Y4260-3900,Zhao-PRL-Y4260-3900-2,Wang-Y4260-No-hadro,WangCPC-Y4390,Mole-GuoFK,
 Mole-GuoFK-DongXK,Mole-YanMJ,Mole-Sundu,Mole-JinHY,Mole-LiuX-chi-rho},  kinematical effects \cite{Chen-He-Liu-4260-1,Chen-He-Liu-4260-2,CC-Effects-1,CC-Effects-2}, baryonium states \cite{Qiao-CF-4260}, etc. The $Y(4260)$, which is the milestone of the $Y$ states,  has been extensively studied.

In the present work, we will focus on the scenario of tetraquark states.  In Ref.\cite{Maiani-4260}, L. Maiani et al assign the $Y(4260)$ as the first orbital excitation
of a (scalar)diquark-(scalar)antidiquark state $[cs][\bar{c}\bar{s}]$  based on the  spin-spin and spin-orbit  interactions.
In Ref.\cite{Ali-Maiani-Y}, A. Ali et al investigate  the hidden-charm P-wave tetraquarks and the newly observed excited charmed $\Omega_c$ states in the diquark model using the  spin-spin, spin-orbit and tensor interactions,  and observe that the preferred  assignments of the ground state tetraquark states with $L=1$ are the $Y(4220)$, $Y(4330)$, $Y(4390)$, $Y(4660)$ rather than the  $Y(4008)$, $Y(4260)$, $Y(4360)$, $Y(4660)$. However, the observation of the process $Y(4260) \to Z_c(3900)^\pm \pi^\mp \to J/\psi \pi^+\pi^-$ disfavors  assigning  the $Y(4230)$ as tetraquark state with the symbolic quark constituents $cs\bar{c}\bar{s}$
\cite{Y-Zc3900-BESIII,Y-Zc3900-Belle}.

In Ref.\cite{Vector-Tetra-WZG-P-wave}, we introduce an explicit P-wave between the diquark and antidiquark to construct the four-quark  currents,  and study the  vector tetraquark states with the QCD sum rules systematically, and obtain the lowest vector tetraquark masses up to now.
  The  predictions support  assigning the
 $Y(4220/4260)$,  $Y(4320/4360)$ and $Y(4390)$  as the vector tetraquark   states with a relative P-wave between the diquark ($qc$) and antidiquark ($\bar{q}\bar{c}$) pair.

In Ref.\cite{WZG-Vector-NPB},  we take the scalar, pseudoscalar, axialvector, vector and tensor  (anti)diquarks as the basic building blocks   to construct  vector and tensor  four-quark currents without introducing explicit P-waves, as the P-waves are implied in negative-parity of the (anti)diquarks, and explore the mass spectrum of the vector hidden-charm tetraquark states via the QCD sum rules comprehensively,  and obtain the lowest tetraquark  mass about $4.35\,\rm{GeV}$ and  revisit the assignments of the  $Y$ states. At the energy about $4.5\,\rm{GeV}$, we obtain three hidden-charm tetraquark states with the $J^{PC}=1^{--}$, the tetraquark states with the symbolic structures  $[uc]_{\tilde{V}}[\overline{dc}]_{A}-[uc]_{A}[\overline{dc}]_{\tilde{V}}$,
 $[uc]_{\tilde{A}}[\overline{dc}]_{V}+[uc]_{V}[\overline{dc}]_{\tilde{A}}$ and
 $[uc]_{S}[\overline{dc}]_{\tilde{V}}-[uc]_{\tilde{V}}[\overline{dc}]_{S}$ have
 the masses $4.53\pm0.07\, \rm{GeV}$, $4.48\pm0.08\,\rm{GeV}$ and $4.50\pm0.09\,\rm{GeV}$, respectively,
  thus we have three candidates for the newly observed $Y(4500)$ \cite{X4500-BESIII}. At the energy about $4.7\,\rm{GeV}$, we obtain the mass $4.69\pm0.08\,\rm{GeV}$ for the $[uc]_{A}[\overline{dc}]_{A}$ tetraquark state with the $J^{PC}=1^{--}$, which is in very good agreement with the newly observed $Y(4708)$ \cite{Y4708-BES}. It is not necessary that the $Y$ states below and above $4.4\,\rm{GeV}$ to have the same structures.

We cannot assign a hadron unambiguously with the mass alone, we have to explore  the decay width to make more robust assignment.
In this work, we tentatively assign the $Y(4230)$ as the tetraquark state $|S_{qc}, S_{\bar{q}\bar{c}}; S, L; J\rangle=|0, 0; 0, 1; 1\rangle$  with the $J^{PC}=1^{--}$ according to the calculations in Ref.\cite{Vector-Tetra-WZG-P-wave},  and   extend our previous works to study the three-body strong decays $Y(4230)\to J/\psi \pi^+\pi^-$,  $J/\psi K^+K^-$,  $\bar{D}^{*-}D^0\pi^+$ and $\bar{D}^{*-}D^{*0}\pi^+$ with the light-cone QCD sum rules \cite{WZG-Y4500-decay}.  In Ref.\cite{WZG-Y4500-decay}, we tentatively assign the $Y(4500)$ as the $[uc]_{\tilde{A}}[\overline{uc}]_{V}+[uc]_{V}[\overline{uc}]_{\tilde{A}}+[dc]_{\tilde{A}}[\overline{dc}]_{V}
+[dc]_{V}[\overline{dc}]_{\tilde{A}}$ tetraquark state with the $J^{PC}=1^{--}$, and suggest  to   calculate the four-meson  coupling constants via the light-cone QCD sum rules directly based on rigorous quark-hadron duality, then study three-body decay $Y(4500)\to \bar{D}^{*-}D^{*0}\pi^+$.

The article is arranged as follows:  we obtain  the light-cone QCD sum rules for the  four-meson coupling constants in section 2; in section 3, we present numerical results and discussions; section 4 is reserved for our conclusion.

\section{Light-cone QCD sum rules for  the  four-meson  coupling constants}
Firstly, we write down  the three-point correlation functions  $\Pi_{\mu\alpha\beta}(p,q)$, $\Pi_{\mu\alpha}^1(p,q)$ and $\Pi_{\mu\alpha}^2(p,q)$  in the light-cone QCD sum rules,
\begin{eqnarray}
\Pi_{\mu\alpha\beta}(p,q)&=&i^2\int d^4xd^4y \, e^{-ip\cdot x}e^{-iq\cdot y}\, \langle 0|T\left\{J_{\mu}^{Y}(0)J_\alpha^{D^{*+}}(x)J^{\bar{D}^{*0}}_{\beta}(y)\right\}|\pi(r)\rangle\, , \nonumber\\
\Pi^1_{\mu\alpha}(p,q)&=&i^2\int d^4xd^4y \, e^{-ip\cdot x}e^{-iq\cdot y}\, \langle 0|T\left\{J_{\mu}^{Y}(0)J_\alpha^{D^{*+}}(x)J^{\bar{D}^{0}}(y)\right\}|\pi(r)\rangle\, , \nonumber\\
\Pi^2_{\mu\alpha}(p,q)&=&i^2\int d^4xd^4y \, e^{-ip\cdot x}e^{-iq\cdot y}\, \langle 0|T\left\{J_{\mu}^{Y}(0)J_\alpha^{J/\psi}(x)J^{\pi^+}(y)\right\}|\pi(r)\rangle\, ,
\end{eqnarray}
where the currents
\begin{eqnarray}\label{current-JY}
J_{\mu}^{Y}(0)&=&\frac{\varepsilon^{ijk}\varepsilon^{imn}}{2}\Big\{u^{T}_j(0)C\gamma_5 c_k(0)\stackrel{\leftrightarrow}{\partial}_\mu \bar{u}_m(0)\gamma_5 C \bar{c}^{T}_n(0)\nonumber\\
&&+d^{T}_j(0)C\gamma_5 c_k(0)\stackrel{\leftrightarrow}{\partial}_\mu \bar{d}_m(0)\gamma_5 C \bar{c}^{T}_n(0) \Big\}\, ,
\end{eqnarray}
\begin{eqnarray}
J_{\alpha}^{D^{*+}}(x)&=&\bar{d}(x)\gamma_{\alpha} c(x) \, ,\nonumber \\
J_{\beta}^{\bar{D}^{*0}}(x)&=&\bar{c}(y)\gamma_{\beta} u(y) \, ,\nonumber \\
J^{\bar{D}^{0}}(y)&=&\bar{c}(y)i\gamma_{5} u(y) \, ,\nonumber \\
J_{\alpha}^{J/\psi}(x)&=&\bar{c}(x)\gamma_{\alpha} c(x) \, ,\nonumber \\
J^{\pi^+}(x)&=&\bar{d}(y)i\gamma_{5} u(y) \, ,
\end{eqnarray}
interpolate the mesons $Y(4230)$,  $\bar{D}^*$, $D^*$, $D$, $J/\psi$ and $\pi$ respectively \cite{Vector-Tetra-WZG-P-wave}, the $|\pi(r)\rangle$ is the external $\pi$ state, the derivative  $\stackrel{\leftrightarrow}{\partial}_\mu=\stackrel{\rightarrow}{\partial}_\mu-\stackrel{\leftarrow}{\partial}_\mu$ embodies  the P-wave effects. We resort to the correlation functions  $\Pi_{\mu\alpha\beta}(p,q)$, $\Pi_{\mu\alpha}^1(p,q)$ and $\Pi_{\mu\alpha}^2(p,q)$ to explore the hadronic coupling constants in the three-body strong decays $Y(4230)\to\bar{D}^* D^* \pi^+$, $\bar{D}^* D \pi^+$ and  $ J/\psi \pi^- \pi^+$, respectively.

At the hadron side, we insert  a complete set of intermediate hadronic states having potential  couplings  with the interpolating currents into the three-point correlation functions $\Pi_{\mu\alpha\beta}(p,q)$, $\Pi_{\mu\alpha}^1(p,q)$ and $\Pi_{\mu\alpha}^2(p,q)$, and  isolate the ground state contributions explicitly,
\begin{eqnarray}\label{Hadron-CT-1}
\Pi_{\mu\alpha\beta}(p,q)&=& \lambda_Y f_{D^*}^2m_{D^*}^2 \frac{-iG_{A}r_\tau+iG_{B}p^\prime_\tau}{(m_{Y}^2-p^{\prime2})(m_{\bar{D}^*}^2-p^2)(m_{D^*}^2-q^2)}\varepsilon^{\rho\sigma\lambda\tau}  \left( -g_{\mu\rho}+\frac{p^\prime_\mu p^\prime_\rho}{p^{\prime2}}\right)\nonumber\\
&&\left( -g_{\alpha\sigma}+\frac{p_\alpha p_\sigma}{p^2}\right)\left( -g_{\lambda\beta}+\frac{q_\lambda q_\beta}{q^2}\right) + \cdots\, ,
\end{eqnarray}
\begin{eqnarray}\label{Hadron-CT-2}
\Pi^1_{\mu\alpha}(p,q)&=& \frac{\lambda_Y f_{D^*}m_{D^*}f_{D}m_{D}^2}{m_c} \frac{iG_{C}}{(m_{Y}^2-p^{\prime2})(m_{\bar{D}^*}^2-p^2)(m_{D}^2-q^2)}  \left( -g_{\mu\rho}+\frac{p^\prime_\mu p^\prime_\rho}{p^{\prime2}}\right)\nonumber\\
&&\left( -g_{\alpha\rho}+\frac{p_\alpha p_\rho}{p^2}\right) + \cdots\, , \nonumber\\
&=&i\Pi_{C}(p^{\prime2},p^2,q^2)\,g_{\mu\alpha}+\cdots\, ,
\end{eqnarray}
\begin{eqnarray}\label{Hadron-CT-3}
\Pi^2_{\mu\alpha}(p,q)&=& \frac{\lambda_Y f_{J/\psi}m_{J/\psi}f_{\pi}m_{\pi}^2}{m_u+m_d} \frac{iG_{D}\,q\cdot r}{(m_{Y}^2-p^{\prime2})(m_{J/\psi}^2-p^2)(m_{\pi}^2-q^2)}  \left( -g_{\mu\rho}+\frac{p^\prime_\mu p^\prime_\rho}{p^{\prime2}}\right)\nonumber\\
&&\left( -g_{\alpha\rho}+\frac{p_\alpha p_\rho}{p^2}\right) + \cdots\, ,\nonumber\\
&=&i\Pi_{D}(p^{\prime2},p^2,q^2)\,q\cdot r\,g_{\mu\alpha}+\cdots\, ,
\end{eqnarray}
where
\begin{eqnarray}
\Pi_{C}(p^{\prime2},p^2,q^2)&=& \frac{\lambda_Y f_{D^*}m_{D^*}f_D m_{D}^2}{m_c} \frac{G_{C}}{(m_{Y}^2-p^{\prime2})(m_{\bar{D}^*}^2-p^2)(m_{D}^2-q^2)}  + \cdots\, ,\nonumber\\
\Pi_{D}(p^{\prime2},p^2,q^2)&=& \frac{\lambda_Y f_{J/\psi}m_{J/\psi}f_\pi m_{\pi}^2}{m_u+m_d} \frac{G_{D}}{(m_{Y}^2-p^{\prime2})(m_{J/\psi}^2-p^2)(m_{\pi}^2-q^2)}  + \cdots\, ,
\end{eqnarray}
$p^\prime=p+q+r$, the  decay constants $\lambda_Y$, $f_{D^*}$, $f_{\bar{D}^*}$, $f_D$, $f_{J/\psi}$, $f_{\pi}$ and hadronic coupling constants $G_A$, $G_B$, $G_C$, $G_D$ are defined by,
\begin{eqnarray}
\langle0|J_{\mu}^{Y}(0)|Y_c(p^\prime)\rangle&=&\lambda_{Y} \varepsilon_{\mu} \,\, , \nonumber \\
\langle0|J_{\alpha}^{D^*\dag}(0)|\bar{D}^*(p)\rangle&=&f_{\bar{D}^*} m_{\bar{D}^*} \xi_\alpha \,\, , \nonumber \\
\langle0|J_{\beta}^{\bar{D}^{*\dag}}(0)|D^*(q)\rangle&=&f_{D^*} m_{D^*} \zeta_\beta \,\, ,\nonumber \\
\langle0|J_{\alpha}^{J/\psi}(0)|J/\psi(p)\rangle&=&f_{J/\psi} m_{J/\psi} \xi_\alpha \,\, , \nonumber \\
\langle0|J^{\bar{D}^{\dag}}(0)|D(q)\rangle&=&\frac{f_{D} m_{D}^2}{m_c}  \,\, ,\nonumber \\
\langle0|J^{\pi^{\dag}}(0)|\pi(q)\rangle&=&\frac{f_{\pi} m_{\pi}^2}{m_u+m_d}  \,\, ,
\end{eqnarray}
\begin{eqnarray}\label{define-G-pi-X}
\langle Y_c(p^\prime)|\bar{D}^*(p)D^*(q)\pi(r)\rangle&=&G_{A}\varepsilon^{\rho\sigma\lambda\tau}\varepsilon^*_\rho\xi_\sigma\zeta_\lambda r_\tau-G_{B}\varepsilon^{\rho\sigma\lambda\tau}\varepsilon^*_\rho\xi_\sigma\zeta_\lambda p^\prime_\tau \,\, ,\nonumber \\
\langle Y_c(p^\prime)|\bar{D}^*(p)D(q)\pi(r)\rangle&=&-G_{C}\,\varepsilon^* \cdot \xi  \,\, ,\nonumber \\
\langle Y_c(p^\prime)|J/\psi(p)\pi(q)\pi(r)\rangle&=&-G_{D}\,q \cdot r\,\varepsilon^* \cdot \xi  \,\, ,
\end{eqnarray}
 the $\varepsilon_{\mu}$, $\xi_\alpha$ and $\zeta_\beta$  are polarization vectors of the $Y(4230)$, $\bar{D}^*$($J/\psi$) and  $D^*$ mesons, respectively. In the correlation functions $\Pi_{\mu\alpha}^1(p,q)$ and $\Pi_{\mu\alpha}^2(p,q)$, see Eqs.\eqref{Hadron-CT-2}-\eqref{Hadron-CT-3}, there are other tensor structures, which lead to different QCD sum rules, those QCD sum rules have shortcomings in one way or the other, and we discard them.

 In the isospin limit, $m_u=m_d$, $f_{D^*}=f_{\bar{D}^*}$ and $m_{D^*}=m_{\bar{D}^*}$. The vertex $\langle Y_c(p^\prime)|J/\psi(p)K(q)K(r)\rangle$ $=-G_{E}\,q \cdot r\,\varepsilon \cdot \xi $,  in the $SU(3)$ limit, $G_{E}=G_{D}$, we have a universal coupling constant.

The tensor structures in the correlation function $\Pi_{\mu\alpha\beta}(p,q)$, see Eq.\eqref{Hadron-CT-1}, are complex, we should simplify them, and  thus facilitate the calculations at the QCD side.
We multiply Eq.\eqref{Hadron-CT-1} with the tensor $\varepsilon_{\theta\omega}{}^{\alpha\beta}$ and obtain
\begin{eqnarray}\label{Hadron-CT-W}
\widetilde{\Pi}_{\mu\theta\omega}(p,q)&=& \varepsilon_{\theta\omega}{}^{\alpha\beta}\,\Pi_{\mu\alpha\beta}(p,q)\nonumber\\
&=&\lambda_Y f_{D^*}^2m_{D^*}^2 \frac{iG_{A}\left(g_{\mu\omega}r_\theta-g_{\mu\theta}r_\omega\right)-iG_{B}
\left(g_{\mu\omega}p^\prime_\theta-g_{\mu\theta}p^\prime_\omega\right)}{(m_{Y}^2-p^{\prime2})(m_{\bar{D}^*}^2-p^2)
(m_{D^*}^2-q^2)}  + \cdots\, , \nonumber\\
&=&\Big\{i\Pi_{A}(p^{\prime2},p^2,q^2)-i\Pi_{B}(p^{\prime2},p^2,q^2)\Big\}
\left(g_{\mu\omega}r_\theta-g_{\mu\theta}r_\omega\right) \nonumber\\
&&-i\Pi_{B}(p^{\prime2},p^2,q^2)
\left(g_{\mu\omega}q_\theta-g_{\mu\theta}q_\omega\right)+\cdots \, ,
\end{eqnarray}
where
\begin{eqnarray}
\Pi_{A}(p^{\prime2},p^2,q^2)&=& \lambda_Y f_{D^*}^2m_{D^*}^2\frac{ G_{A}}{(m_{Y}^2-p^{\prime2})(m_{\bar{D}^*}^2-p^2) (m_{D^*}^2-q^2)}  + \cdots\, ,\nonumber\\
\Pi_{B}(p^{\prime2},p^2,q^2)&=& \lambda_Y f_{D^*}^2m_{D^*}^2 \frac{ G_{B}}{(m_{Y}^2-p^{\prime2})(m_{\bar{D}^*}^2-p^2)(m_{D^*}^2-q^2)}  + \cdots\, ,
\end{eqnarray}
then we choose the tensor structures $g_{\mu\omega}r_\theta-g_{\mu\theta}r_\omega$ and $g_{\mu\omega}q_\theta-g_{\mu\theta}q_\omega$ to explore the   $G_{A}$ and $G_{B}$, respectively. Here we also neglect the tensor structures, which cannot end up with good QCD sum rules.

After choosing the best tensor structures therefore the best components of the correlation functions, we obtain the hadronic  spectral densities $\rho_H(s^\prime,s,u)$ through triple  dispersion relation,
\begin{eqnarray}
\Pi_{H}(p^{\prime2},p^2,q^2)&=&\int_{\Delta_s^{\prime2}}^\infty ds^{\prime} \int_{\Delta_s^2}^\infty ds \int_{\Delta_u^2}^\infty du \frac{\rho_{H}(s^\prime,s,u)}{(s^\prime-p^{\prime2})(s-p^2)(u-q^2)}\, ,
\end{eqnarray}
where the $\Delta_{s}^{\prime2}$, $\Delta_{s}^{2}$ and
$\Delta_{u}^{2}$ are the thresholds, we add the subscript $H$ to denote the hadron side, and $H=A$, $B$, $C$ and $D$.

Now we explore the QCD side with an eye on the relevant tensor structures, we carry out   the operator product expansion up to the vacuum condensates of dimension 5 and neglect the tiny gluon condensate contributions \cite{WZG-Y4500-decay,WZG-ZJX-Zc-Decay,WZG-Y4660-Decay}, and choose  the $\pi$-meson  light-cone wave functions \cite{PBall-LCDF}, which are defined by
\begin{eqnarray}
\langle 0|\bar{d}(0)\gamma_\mu\gamma_5 u(x)|\pi(r)\rangle &=&if_\pi r_\mu \int_0^1 du e^{-iur\cdot x} \varphi_{\pi}(u)+\cdots\, , \nonumber\\
\langle 0|\bar{d}(0)\sigma_{\mu\nu}\gamma_5 u(x)|\pi(r)\rangle &=&\frac{i}{6}\frac{f_\pi m_\pi^2}{m_u+m_d} \left(r_\mu x_\nu -r_\nu x_\mu \right) \int_0^1 du e^{-iur\cdot x} \varphi_{\sigma}(u) \, , \nonumber\\
\langle 0|\bar{d}(0)i\gamma_5 u(x)|\pi(r)\rangle &=& \frac{f_\pi m_\pi^2}{m_u+m_d}  \int_0^1 du e^{-iur\cdot x} \varphi_{5}(u) \, ,
\end{eqnarray}
and take the approximation,
\begin{eqnarray} \label{qGq}
\langle 0|\bar{d}(x_1)\sigma_{\mu\nu}\gamma_5g_sG_{\alpha\beta}(x_2) u(x_3)|\pi(r)\rangle &=&if_{3\pi}\left( r_\mu r_\alpha g_{\nu\beta}+r_\nu r_\beta g_{\mu\alpha}-r_\nu r_\alpha g_{\mu\beta}-r_\mu r_\beta g_{\nu\alpha}\right) \, ,
\end{eqnarray}
for the twist-3 quark-gluon light-cone wave  functions.
In calculations, we find that the terms proportional to $m_\pi^2$  are greatly suppressed and neglect them safely, except for the case that the terms proportional to $\frac{m_\pi^2}{m_u+m_d}$ are Chiral enhanced due to the Gell-Mann-Oakes-Renner relation $\frac{f_\pi m_\pi^2}{m_u+m_d}=-\frac{2\langle\bar{q}q\rangle}{f_\pi}$, and we take account of their contributions  fully. Now we list out the $\pi$-meson light-cone wave  functions of twist-2 and twist-3 explicitly,
\begin{eqnarray}
\varphi_\pi(u)&=&6u\bar{u}\left[1+A_2\frac{3}{2}\left(5t^2-1 \right)+A_4 \frac{15}{8}\left(21t^4-14t^2+1 \right)  \right]\, ,\nonumber\\
\varphi_5(u)&=&1+B_2\frac{1}{2}\left(3t^2-1 \right)+B_4 \frac{1}{8}\left(35t^4-30t^2+3 \right)  \, ,\nonumber\\
\varphi_\sigma(u)&=&6u\bar{u}\left[1+C_2\frac{3}{2}\left(5t^2-1 \right) \right]\, ,
\end{eqnarray}
where $t=2u-1$, and the coefficients $A_2=0.44$, $A_4=0.25$, $B_2=0.43$, $B_4=0.10$, $C_2=0.09$, and the decay constant $f_{3\pi}=0.0035\,\rm{GeV}^2$ at the energy scale $\mu=1\,\rm{GeV}$ \cite{PBall-LCDF,Braun-f3pi}.
We neglect the twist-4 light-cone wave functions due to their small contributions considering the associated factor $m_\pi^2$.

In the soft limit $r_\mu \to 0$, $\tilde{q}^2=(q+r)^2=q^2$ and  $\tilde{p}^2=(p+r)^2=p^2$, we set $\Pi_{A/B/C/D}(p^2,\tilde{p}^{2},q^2)$ or $\Pi_{A/B/C/D}(p^2,\tilde{q}^{2},q^2)=\Pi_{A/B/C/D}(p^2,q^2)$ to simplify the analytical expressions, then we obtain the QCD spectral densities $\rho_{QCD}(s,u)$  through double dispersion relation,
\begin{eqnarray}
\Pi^{QCD}_{A/B/C/D}(p^2,q^2)&=& \int_{\Delta_s^2}^\infty ds \int_{\Delta_u^2}^\infty du \frac{\rho_{QCD}(s,u)}{(s-p^2)(u-q^2)}\, ,
\end{eqnarray}
again the $\Delta_s^2$ and $\Delta_u^2$  are the thresholds, we add the superscript or subscript $QCD$ to denote   the QCD side.

We match the hadron side with the QCD side  bellow the continuum thresholds $s_0$ and $u_0$ to obtain rigorous quark-hadron  duality  \cite{WZG-ZJX-Zc-Decay,WZG-Y4660-Decay},
 \begin{eqnarray}
  \int_{\Delta_s^2}^{s_{0}}ds \int_{\Delta_u^2}^{u_0}du  \frac{\rho_{QCD}(s,u)}{(s-p^2)(u-q^2)}&=& \int_{\Delta_s^2}^{s_0}ds \int_{\Delta_u^2}^{u_0}du  \left[ \int_{\Delta_{s}^{\prime2}}^{\infty}ds^\prime  \frac{\rho_H(s^\prime,s,u)}{(s^\prime-p^{\prime2})(s-p^2)(u-q^2)} \right]\, .
\end{eqnarray}
To facilitate the procedure beyond formal  calculations,  we carry out  the integral over $ds^\prime$ firstly,
then write down the hadron representation explicitly,
\begin{eqnarray}
\Pi_{A/B}(p^{\prime2},p^2,q^2)&=& \frac{\lambda_Y f_{D^*}^2m_{D^*}^2 \, G_{A/B}}{(m_{Y}^2-p^{\prime2})(m_{\bar{D}^*}^2-p^2)
(m_{D^*}^2-q^2)}  +\frac{C_{A/B}}{(m_{\bar{D}^*}^2-p^2)(m_{D^*}^2-q^2)}+\cdots\, , \nonumber\\
\Pi_{C}(p^{\prime2},p^2,q^2)&=& \frac{\lambda_Y f_{D^*}m_{D^*}f_{D}m_{D}^2 \, G_{C}}{m_c(m_{Y}^2-p^{\prime2})(m_{\bar{D}^*}^2-p^2)
(m_{D}^2-q^2)}  +\frac{C_{C}}{(m_{\bar{D}^*}^2-p^2)(m_{D}^2-q^2)}+\cdots\, , \nonumber\\
\Pi_{D}(p^{\prime2},p^2,q^2)&=& \frac{\lambda_Y f_{J/\psi}m_{J/\psi}\mu_{\pi} \, G_{D}}{(m_{Y}^2-p^{\prime2})(m_{J/\psi}^2-p^2)
(m_{\pi}^2-q^2)}  +\frac{C_{D}}{(m_{J/\psi}^2-p^2)(m_{\pi}^2-q^2)}+\cdots\, ,
\end{eqnarray}
where $\mu_\pi=\frac{f_\pi m_\pi^2}{m_u+m_d}$, we introduce the parameters $C_{A/B/C/D}$ to parameterize the contributions involving   the higher resonances and continuum states in the $s^\prime$ channel,
\begin{eqnarray}
C_{A/B}&=&\int_{s^\prime_0}^{\infty}ds^\prime\frac{\tilde{\rho}_{A/B}(s^\prime,m_{\bar{D}^*}^2,m_{D^*}^2)}{
s^\prime-p^{\prime2}}\, , \nonumber\\
C_{C}&=&\int_{s^\prime_0}^{\infty}ds^\prime\frac{\tilde{\rho}_{C}(s^\prime,m_{\bar{D}^*}^2,m_{D}^2)}{
s^\prime-p^{\prime2}}\, , \nonumber\\
C_{D}&=&\int_{s^\prime_0}^{\infty}ds^\prime\frac{\tilde{\rho}_{D}(s^\prime,m_{J/\psi}^2,m_{\pi}^2)}{
s^\prime-p^{\prime2}}\, ,
\end{eqnarray}
where the hadronic spectral densities $\rho_{H}(s^\prime,s,u)=\tilde{\rho}_{A/B}(s^\prime,s,u)\delta(s-m_{\bar{D}^*}^2)\delta(u-m_{D^*}^2)$,
$\tilde{\rho}_{C}(s^\prime,s,u)\delta(s-m_{\bar{D}^*}^2)\delta(u-m_{D}^2)$ and $\tilde{\rho}_{D}(s^\prime,s,u)\delta(s-m_{J/\psi}^2)\delta(u-m_{\pi}^2)$, respectively. We have no knowledge about the hadronic interactions so as to obtain the analytical expressions of the spectral densities  $\tilde{\rho}_{A/B/C/D}(s^\prime,s,u)$ at the region $s^\prime> s^\prime_0$,
fortunately, we take account of their contributions for the first time. In previous works except for ours, they are neglected without proving feasibility.

In numerical calculations, we  take the unknown functions $C_{A/B/C/D}$ as free parameters and adjust the values to obtain flat platforms  for the hadronic coupling constants $G_{A/B/C/D}$ in regard to variations of the Borel parameters. Such a method works well in the case of three-hadron  contact vertexes \cite{WZG-ZJX-Zc-Decay,WZG-Y4660-Decay,WZG-X4140-decay,
WZG-X4274-decay,WZG-Z4600-decay,WZG-Zcs3985-decay,WZG-Zcs4123-decay}, and four-hadron  contact vertexes $Y(4500)\bar{D}^*D^*\pi$ \cite{WZG-Y4500-decay}.
We set $p^{\prime2}=p^2$ in the correlation functions $\Pi_H(p^{\prime 2},p^2,q^2)$ for simplicity, and  perform  double Borel transform with respect  to  $P^2=-p^2$ and $Q^2=-q^2$ respectively, then we set $T_1^2=T_2^2=T^2$  to obtain  four QCD sum rules,

\begin{eqnarray} \label{pi-SR-A}
&&\frac{\lambda_{YD^*D^*}G_{A}}{m_{Y}^2-m_{\bar{D}^*}^2} \left[ \exp\left(-\frac{m_{\bar{D}^*}^2}{T^2} \right)-\exp\left(-\frac{m_{Y}^2}{T^2} \right)\right]\exp\left(-\frac{m_{D^*}^2}{T^2} \right)+C_{A} \exp\left(-\frac{m_{\bar{D}^*}^2+m_{D^*}^2}{T^2}  \right) \nonumber\\
&&= \frac{\mu_\pi m_c}{576\pi^2}\int_{m_c^2}^{s_{D^*}^0}ds\int_{0}^{1}du\,\varphi_{\sigma}(u)\left(1-\frac{m_c^2}{s}\right)^3
\left(s+m_c^2+\frac{s^2-m_c^4}{T^2}\right)\exp\left(-\frac{s+m_c^2+u\bar{u}m_\pi^2}{T^2}\right)\nonumber\\
&&+\frac{f_\pi}{192\pi^2}\int_{m_c^2}^{s_{D^*}^0}ds\int_{0}^{1}du\,\varphi_{\pi}(u)
\left(1-\frac{m_c^2}{s}\right)^3\left(s^2+8sm_b^2-m_b^4\right)\exp\left(-\frac{s+m_c^2+u\bar{u}m_\pi^2}{T^2}\right)\nonumber\\
&&-\frac{\mu_\pi m_c}{24\pi^2}\int_{m_c^2}^{s_{D^*}^0}ds\int_{0}^{1}du\,\varphi_{5}(u)\,us\left(1-\frac{m_c^2}{s}\right)^3\exp\left(-\frac{s+m_c^2+u\bar{u}m_\pi^2}{T^2}\right)\nonumber\\
&&+\frac{\mu_\pi m_c}{288\pi^2}\int_{m_c^2}^{s_{D^*}^0}ds\int_{0}^{1}du\,\varphi_{\sigma}(u)\left(1-\frac{m_c^2}{s}\right)^2
\left(-s+7m_c^2\right)\exp\left(-\frac{s+m_c^2+u\bar{u}m_\pi^2}{T^2}\right)\nonumber\\
&&+\frac{\mu_{\pi}m_c^2}{9}\left[\langle\bar{q}q\rangle-\frac{m_c^2\langle\bar{q}g_s\sigma Gq\rangle}{4T^4}\right]\int_{0}^{1}du\,\varphi_{\sigma}(u)\exp\left(-\frac{2m_c^2+u\bar{u}m_\pi^2}{T^2}\right)\nonumber\\
&&-\frac{\mu_{\pi}\langle\bar{q}g_s\sigma Gq\rangle}{24} \int_{0}^{1}du\,\varphi_{5}(u)\exp\left(-\frac{2m_c^2+u\bar{u}m_\pi^2}{T^2}\right)\nonumber\\
&&+\frac{\mu_{\pi}\langle\bar{q}g_s\sigma Gq\rangle}{108}\int_{0}^{1}du\,\varphi_{\sigma}(u)\left(-2+\frac{m_c^2}{T^2}\right)\exp\left(-\frac{2m_c^2+u\bar{u}}{T^2}\right)\, ,
\end{eqnarray}

\begin{eqnarray} \label{pi-SR-B}
&&\frac{\lambda_{YD^*D^*}G_{B}}{m_{Y}^2-m_{\bar{D}^*}^2} \left[ \exp\left(-\frac{m_{\bar{D}^*}^2}{T^2} \right)-\exp\left(-\frac{m_{Y}^2}{T^2} \right)\right]\exp\left(-\frac{m_{D^*}^2}{T^2} \right)+C_{B} \exp\left(-\frac{m_{\bar{D}^*}^2+m_{D^*}^2}{T^2}  \right) \nonumber\\
&&= -\frac{\mu_{\pi}\langle\bar{q}g_s\sigma Gq\rangle}{24}\int_{0}^{1}du\varphi_{5}(u)\exp\left(-\frac{2m_c^2+u\bar{u}m_\pi^2}{T^2}\right) \, ,
\end{eqnarray}

\begin{eqnarray} \label{pi-SR-C}
&&\frac{\lambda_{YD^*D}G_{C}}{m_{Y}^2-m_{\bar{D}^*}^2} \left[ \exp\left(-\frac{m_{\bar{D}^*}^2}{T^2} \right)-\exp\left(-\frac{m_{Y}^2}{T^2} \right)\right]\exp\left(-\frac{m_{D}^2}{T^2} \right)+C_{C} \exp\left(-\frac{m_{\bar{D}^*}^2+m_{D}^2}{T^2}  \right) \nonumber\\
&&=\frac{\mu_\pi}{192\pi^2}\int_{m_c^2}^{s_D^0}ds\int_{0}^{1}du\,\varphi_{5}(u)
\left(1-\frac{m_c^2}{s}\right)^3s(s+2m_c^2)\exp\left(-\frac{s+m_c^2+u\bar{u}m_\pi^2}{T^2}\right)\nonumber\\
&&+\frac{\mu_{\pi}m_c^2}{192\pi^2}\int_{m_c^2}^{s_{D^*}^0}ds\int_{0}^{1}du\,\varphi_{5}(u)
\left(1-\frac{m_c^2}{s}\right)^3\left(2s+m_c^2\right)\exp\left(-\frac{s+m_c^2+u\bar{u}m_\pi^2}{T^2}\right)\nonumber\\
&&+\frac{\mu_{\pi}m_c\langle\bar{q}g_s\sigma Gq\rangle}{24}\int_{0}^{1}du\,\varphi_{5}(u)\exp\left(-\frac{2m_c^2+u\bar{u}m_\pi^2}{T^2}\right) \, ,
\end{eqnarray}

\begin{eqnarray} \label{pi-SR-D}
&&\frac{\lambda_{YJ/\psi \pi}G_{D}}{m_{Y}^2-m_{J/\psi}^2} \left[ \exp\left(-\frac{m_{J/\psi}^2}{T^2} \right)-\exp\left(-\frac{m_{Y}^2}{T^2} \right)\right]\exp\left(-\frac{m_{\pi}^2}{T^2} \right)+C_{D} \exp\left(-\frac{m_{J/\psi}^2+m_{\pi}^2}{T^2}  \right) \nonumber\\
&&=\frac{\mu_{\pi}}{72\pi^2} \int_{4m_c^2}^{s^0_{J/\psi}}ds\sqrt{1-\frac{4m_c^2}{s}}\int_{0}^{1}du\,\varphi_{\sigma}(u)(s+2m_c^2)\exp\left(-\frac{s+u\bar{u}m_\pi^2}{T^2}\right)\nonumber\\
&&+\frac{f_\pi m_b}{24\pi^2}\int_{4m_c^2}^{s^0_{J/\psi}}ds\sqrt{1-\frac{4m_c^2}{s}}\int_{0}^{1}du\,\varphi_{\pi}(u)(s-4m_c^2)\exp\left(-\frac{s+u\bar{u}m_\pi^2}{T^2}\right)  \, ,
\end{eqnarray}
where $\lambda_{YD^*D^*}=\lambda_{Y}f_{D^*}^2m^2_{D^*}$, $\lambda_{YD^*D}=\frac{\lambda_{Y}f_{D^*}m_{D^*}f_{D}m_{D}^2}{m_c}$ and $\lambda_{YJ/\psi \pi}=\frac{\lambda_{Y}f_{J/\psi}m_{J/\psi}f_{\pi}m_{\pi}^2}{m_u+m_d}$.

\section{Numerical results and discussions}
We take  the standard values of the vacuum condensates,
$\langle
\bar{q}q \rangle=-(0.24\pm 0.01\, \rm{GeV})^3$,
$\langle\bar{q}g_s\sigma G q \rangle=m_0^2\langle \bar{q}q \rangle$,
$m_0^2=(0.8 \pm 0.1)\,\rm{GeV}^2$     at the   energy scale  $\mu=1\, \rm{GeV}$
\cite{SVZ79,Reinders85,Colangelo-Review},  and take the $\overline{MS}$  mass $m_{c}(m_c)=(1.275\pm0.025)\,\rm{GeV}$ from the Particle Data Group \cite{PDG}. We set $m_u=m_d=0$ and take account of
the energy-scale dependence of  the input parameters,
\begin{eqnarray}
\langle\bar{q}q \rangle(\mu)&=&\langle\bar{q}q \rangle({\rm 1GeV})\left[\frac{\alpha_{s}({\rm 1GeV})}{\alpha_{s}(\mu)}\right]^{\frac{12}{33-2n_f}}\, , \nonumber\\
 \langle\bar{q}g_s \sigma Gq \rangle(\mu)&=&\langle\bar{q}g_s \sigma Gq \rangle({\rm 1GeV})\left[\frac{\alpha_{s}({\rm 1GeV})}{\alpha_{s}(\mu)}\right]^{\frac{2}{33-2n_f}}\, , \nonumber\\
 m_c(\mu)&=&m_c(m_c)\left[\frac{\alpha_{s}(\mu)}{\alpha_{s}(m_c)}\right]^{\frac{12}{33-2n_f}} \, ,\nonumber\\
 \alpha_s(\mu)&=&\frac{1}{b_0t}\left[1-\frac{b_1}{b_0^2}\frac{\log t}{t} +\frac{b_1^2(\log^2{t}-\log{t}-1)+b_0b_2}{b_0^4t^2}\right]\, ,
\end{eqnarray}
  where   $t=\log \frac{\mu^2}{\Lambda_{QCD}^2}$, $b_0=\frac{33-2n_f}{12\pi}$, $b_1=\frac{153-19n_f}{24\pi^2}$, $b_2=\frac{2857-\frac{5033}{9}n_f+\frac{325}{27}n_f^2}{128\pi^3}$,  $\Lambda_{QCD}=210\,\rm{MeV}$, $292\,\rm{MeV}$  and  $332\,\rm{MeV}$ for the flavors  $n_f=5$, $4$ and $3$, respectively  \cite{PDG,Narison-mix}, and we choose  $n_f=4$, and evolve all the input parameters to the typical energy scale  $\mu=1\,\rm{GeV}$.

 At the hadron side, we take $m_{\pi}=0.13957\,\rm{GeV}$, $m_{J/\psi}=3.0969\,\rm{GeV}$,  $f_{\pi}=0.130\,\rm{GeV}$ from the Particle Data Group \cite{PDG},
 $m_{D^*}=2.01\,\rm{GeV}$, $m_{D}=1.87\,\rm{GeV}$, $f_{D^*}=263\,\rm{MeV}$, $f_{D}=208\,\rm{MeV}$, $s^0_{D^*}=6.4\,\rm{GeV}^2$, $s^0_{D}=6.2\,\rm{GeV}^2$   \cite{WangJHEP}, $f_{J/\psi}=0.418 \,\rm{GeV}$  \cite{Becirevic},
 $m_{Y}=4.24\,\rm{GeV}$,   $\lambda_{Y}=2.31 \times 10^{-2}\,\rm{GeV}^6$ \cite{Vector-Tetra-WZG-P-wave} from the QCD sum rules, and  $f_{\pi}m^2_{\pi}/(m_u+m_d)=-2\langle \bar{q}q\rangle/f_{\pi}$ from the Gell-Mann-Oakes-Renner relation.

\begin{figure}
\centering
\includegraphics[totalheight=5cm,width=7cm]{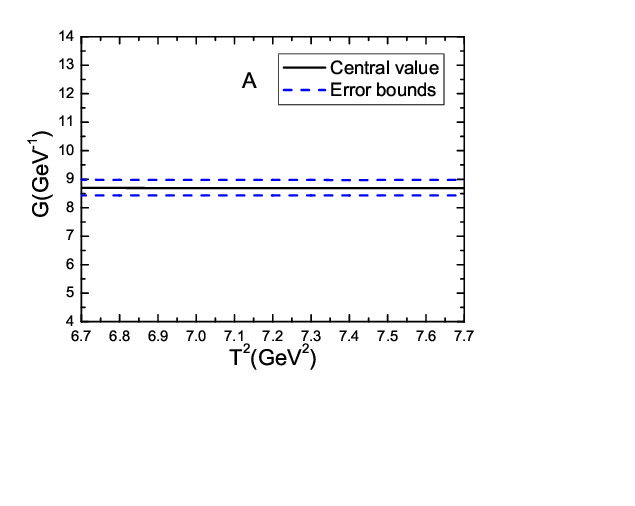}
\includegraphics[totalheight=5cm,width=7cm]{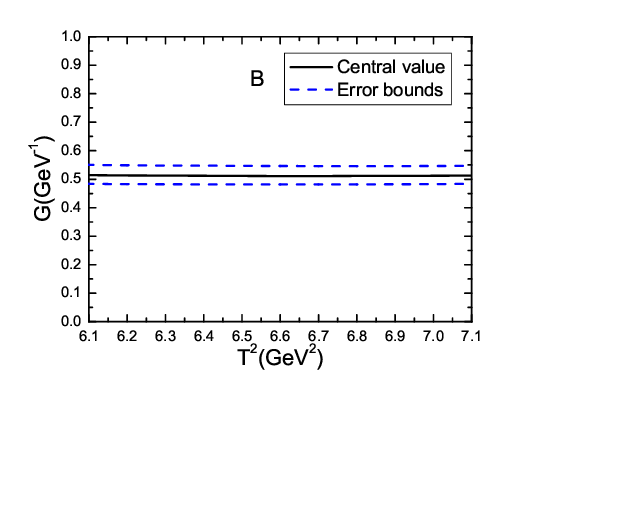}
\includegraphics[totalheight=5cm,width=7cm]{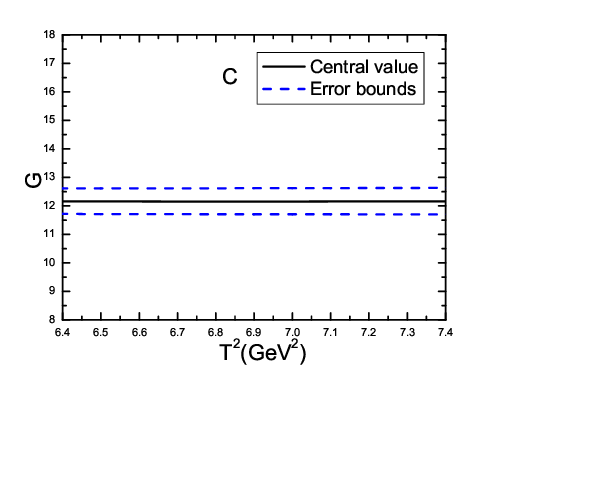}
\includegraphics[totalheight=5cm,width=7cm]{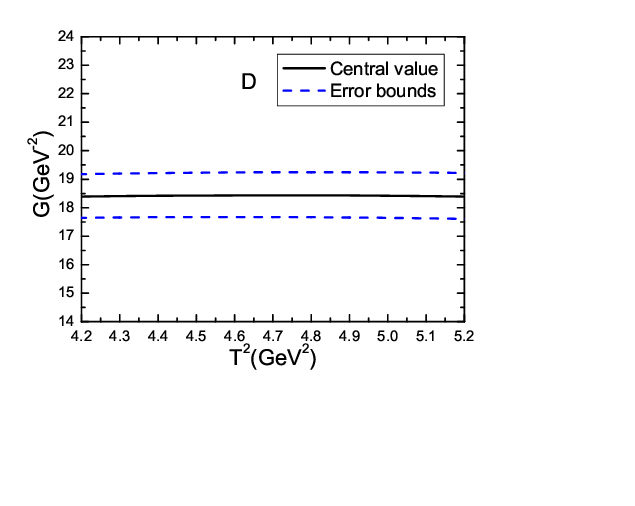}
  \caption{ The hadronic coupling constants with variations of the Borel parameters $T^2$, where the $A$, $B$, $C$ and $D$ denote the $G_{A}$, $G_{B}$, $G_{C}$ and $G_{D}$, respectively.  }\label{G-pi-X}
\end{figure}

In calculations, we fit the free parameters to be $C_{A}=0.00011(T^2-4\,\rm{GeV}^2)\,\rm{GeV}^5$ and
$C_{B}=-0.0000084(T^2-4\,\rm{GeV}^2)\,\rm{GeV}^5$, $C_{C}=0.000179(T^2-4\,\rm{GeV}^2)\,\rm{GeV}^6$, and $C_{D}=0.0015(T^2-4\,\rm{GeV}^2)\,\rm{GeV}^4$
  to obtain uniform flat Borel platforms  $T^2_{max}-T^2_{min}=1\,\rm{GeV}^2$ (just like in our previous works \cite{WZG-Y4500-decay,WZG-ZJX-Zc-Decay,WZG-Y4660-Decay,WZG-X4140-decay,
WZG-X4274-decay,WZG-Z4600-decay,WZG-Zcs3985-decay,WZG-Zcs4123-decay}) via trial and error, where the max and min denote the maximum and minimum values, respectively.
The Borel windows  are $T^2_{A}=(6.7-7.7)\,\rm{GeV}^2$, $T^2_{B}=(6.1-7.1)\,\rm{GeV}^2$,
$T^2_{C}=(6.4-7.4)\,\rm{GeV}^2$ and $T^2_{D}=(4.2-5.2)\,\rm{GeV}^2$,
where  the subscripts $A$, $B$, $C$ and $D$ denote the corresponding QCD  sum rules,  the uncertainties  $\delta G_{A/B/C/D}$ come from the Borel parameters $T^2$ are less than $0.01\, (\rm GeV^{-1}/GeV/GeV^{-2})$. In Fig.\ref{G-pi-X}, we plot the $G_{A}$, $G_{B}$, $G_{C}$ and $G_{D}$ in regard to variations of the Borel parameters. In the Borel windows, there appear very flat platforms  in all channels indeed, it is reasonable and reliable to extract the hadron coupling constants.

If we take  the symbol  $\xi$ to stand for the input parameters at the QCD side, generally speaking, all the uncertainties originate from the QCD parameters,  then  the uncertainties   $\bar{\xi} \to \bar{\xi} +\delta \xi$ result in the uncertainties $\bar{\lambda}_{Y}\bar{f}_{D^*}\bar{f}_{\bar{D}^*}\bar{G}_{A/B} \to \bar{\lambda}_{Y}\bar{f}_{D^*}\bar{f}_{\bar{D}^*}\bar{G}_{A/B}
+\delta\,\bar{\lambda}_{Y}\bar{f}_{D^*}\bar{f}_{\bar{D}^*}\bar{G}_{A/B}$, $\bar{C}_{A/B} \to \bar{C}_{A/B}+\delta C_{A/B}$,
\begin{eqnarray}\label{Uncertainty-4}
\delta\,\bar{\lambda}_{Y}\bar{f}_{D^*}\bar{f}_{\bar{D}^*}\bar{G}_{A/B} &=&\bar{\lambda}_{Y}\bar{f}_{D^*}\bar{f}_{\bar{D}^*}\bar{G}_{A/B}\left( \frac{\delta f_{D^*}}{\bar{f}_{D^*}} +\frac{\delta f_{\bar{D}^*}}{\bar{f}_{\bar{D}^*}}+\frac{\delta \lambda_{Y}}{\bar{\lambda}_{Y}}+\frac{\delta G_{A/B}}{\bar{G}_{A/B}}\right)\, ,
\end{eqnarray}
where  the short overline \,$\bar{}$\, on all the  parameters   denotes the central values.
Direct  calculations indicate that we can set $\delta C_{A/B}=0$ and $\frac{\delta f_{D^*}}{\bar{f}_{D^*}} =\frac{\delta f_{\bar{D}^*}}{\bar{f}_{\bar{D}^*}}=\frac{\delta \lambda_{Y}}{\bar{\lambda}_{Y}}=\frac{\delta G_{A/B}}{\bar{G}_{A/B}}$ approximately.  And the hadronic coupling constants $G_C$ and $G_{D}$ are treated in the same way. In fact, not in all QCD sum rules we can set $\delta C=0$ approximately, if such situations occur, we have to take account of the uncertainties $\delta C$. Now we obtain the hadronic coupling constants routinely,
\begin{eqnarray} \label{HCC-values}
G_{A} &=&8.69 \pm 0.27\,\rm{GeV}^{-1}\, , \nonumber\\
G_{B} &=&0.51\pm 0.03\,\rm{GeV}^{-1}\, ,\nonumber\\
G_{C} &=&12.15 \pm 0.45 \, , \nonumber\\
G_{D} &=&18.42\pm 0.82\,\rm{GeV}^{-2}\, .
\end{eqnarray}

Then we calculate the partial decay widths by taking the hadron masses
$m_{Y}=4.2225\,\rm{GeV}$, $m_{D^{*-}} = 2.01026\,\rm{GeV}$, $m_{D^{*0}}= 2.00685\,\rm{GeV}$, $m_{D^{0}}= 1.86484\,\rm{GeV}$,
$m_{J/\psi}= 3.09690\,\rm{GeV}$, and
$m_{\pi} = 0.13957\,\rm{GeV}$, $m_{K} = 0.493677\,\rm{GeV}$ from the Particle Data Group \cite{PDG},
\begin{eqnarray} \label{Partial-with}
\Gamma\left(Y(4230)\to\bar{D}^* D^* \pi^+\right)&=&0.068^{+0.017}_{-0.015}\,\rm{KeV}\, , \nonumber\\
\Gamma\left(Y(4230)\to\bar{D}^* D \pi^+\right)&=&0.12^{+0.01}_{-0.01}\,\rm{MeV}\, , \nonumber\\
\Gamma\left(Y(4230)\to J/\psi \pi^- \pi^+\right)&=&0.31^{+0.03}_{-0.03}\,\rm{MeV}\, , \nonumber\\
\Gamma\left(Y(4230)\to J/\psi K^- K^+\right)&=&0.013^{+0.001}_{-0.001}\,\rm{MeV}\, ,
\end{eqnarray}
where we set $G_{E}=G_{D}$ in the light flavor $SU(3)$ limit.

The partial decay widths  are much smaller than the average total width  $\Gamma=48 \pm 8 \,\rm{MeV}$ from the Particle Data Group \cite{PDG}, it is obvious that the contact four-meson coupling constants lead to too small partial decay widths, and disfavors observations of the $Y(4230)$ in the three-meson final states,  however, the decays
  $Y(4230) \to \pi^+\pi^- J/\psi$   \cite{BaBar4260-0506},    $ \pi^+\pi^- h_c$ \cite{BES-Y4390}, $ \pi^+D^0D^{*-}$  \cite{BESIII-DDvpi}, $ K^+K^-J/\psi$ \cite{BESIII-KK-4500},  $ D^{*-}D^{*0}\pi^+$  \cite{X4500-BESIII} have been observed experimentally. We expect those decays take place through an intermediate meson,
  \begin{eqnarray}
  Y(4230)&\to&Z_c(3900/4020)^\pm \pi^\mp \to J/\psi \pi^+\pi^-\, ,\,\pi^+\pi^- h_c\, ,\,\pi^\pm (D\bar{D}^{*})^\mp\, ,\,\pi^\pm (D^*\bar{D}^{*})^\mp\, , \nonumber \\
  Y(4230)&\to&J/\psi f_0(500) \to J/\psi \pi^+\pi^- \, , \nonumber \\
  Y(4230)&\to&\bar{D}^{*-} D^{*+}_{0}(2300) \to \bar{D}^{*-}D^0\pi^+ \, , \nonumber \\
  Y(4230)&\to&Z_{cs}(3985)^\pm K^\mp \to J/\psi K^+K^-\, ,
  \end{eqnarray}
  we can search for the intermediate states and precisely measure the branching fractions, which maybe shed light on the nature of the $Y$ states. In fact, the processes  $Y(4230) \to Z_c(3900)^\pm \pi^\mp \to J/\psi \pi^+\pi^-$ have been observed
\cite{Y-Zc3900-BESIII,Y-Zc3900-Belle}. We naively expect that the main decay channels of the vector tetraquark states are two-body strong decays $Y\to D\bar{D}$, $D^*\bar{D}^*$, $D\bar{D}^*$, $D^*\bar{D}$, as they would take place through the Okubo-Zweig-Iizuka super-allowed fall-apart mechanism.

If we assign the  $Y(4230)$ as a $\chi_{c0}\rho^0$ molecule, it is easy to interpret why the decay $Y(4230)\rightarrow
\pi^+\pi^-J/\psi$ has a larger branching fraction  than the decay $Y(4230)\rightarrow D\bar
D$, which has not been observed yet \cite{Mole-LiuX-chi-rho}. Furthermore,  it is a direct consequence that the  decay mode $Y(4230)\rightarrow \pi^+\pi^-J/\psi$
 is more favorable than the  mode $Y(4230)\rightarrow K\bar KJ/\psi$. However, it is difficult (not impossible) to interpret observation of the $Y(4230)$ in the $\bar{D}^* D^* \pi^+$ or $\bar{D}^* D \pi^+$ or $Z_c^\pm(3900) \pi^\mp $ invariant mass spectrum.

 In Ref.\cite{Mole-GuoFK}, Chen et al study the dipion invariant mass spectrum of the  $e^+ e^-
\to Y(4230) \to J/\psi \pi^+\pi^-$ process and the ratio of the cross sections
${\sigma(e^+e^- \to J/\psi K^+ K^-)}/{\sigma(e^+e^- \to J/\psi \pi^+\pi^-)}$, and observe
 that the $SU(3)$ octet state plays a significant
role in those transitions,   the $Y(4230)$ is neither a hybrid nor a conventional charmonium state,
but has a sizeable $\bar{D} D_1$ component, which, however, is not completely dominant. On the other hand, the calculations based on the QCD sum rules indicate that we cannot obtain a $\bar{D} D_1$ molecular state having the mass as low as $4.2\,\rm{GeV}$ \cite{WangCPC-Y4390}. The situation is very complex as there maybe exist mixing effects \cite{Mole-JinHY}.

Recently, the BESIII collaboration measured the Born cross sections of the process $e^{+}e^{-}\rightarrow D_{s}^{\ast}\bar{D}_{s}^{\ast}$  at center-of-mass energies from threshold to $4.95\, \rm{GeV}$ with high precision for the first time, and observed two resonance structures around 4.2 and 4.4 GeV, respectively \cite{Y4790-BES}. The fitted  Breit-Wigner masses are $4186.5\pm9.0\pm3\,\rm{MeV}$ and $4414.5\pm3.2\pm6.0\,\rm{MeV}$, the widths are $55\pm17\pm53\,\rm{MeV}$ and $122.6\pm7.0\pm8.2\,\rm{MeV}$. If they are not the $\psi(4160)$ and $\psi(4415)$,
there are some contributions from the $Y(4230)$, which indicates that the $Y(4230)$
couples more strongly to the $D_{s}^{\ast}\bar{D}_{s}^{\ast}$ mode than to the
modes with charmonium states, as the cross section
of the process  $e^{+}e^{-}\rightarrow D_{s}^{\ast}\bar{D}_{s}^{\ast}$  at 4.23 GeV is roughly one order of
magnitude higher than that of the process  $e^{+}e^{-}\rightarrow J/\psi \pi^+\pi^-$, then the $Y(4230)$ should have some $\bar{c}c\bar{s}s$ components at least. So exploring the processes $Y(4230)\to D\bar{D}$, $D^*\bar{D}^*$, $D\bar{D}^*$, $D^*\bar{D}$ is of great importance.

\section{Conclusion}
 In our previous works, we have proven that a vector tetraquark configuration with a relative P-wave between the scalar diquark pair could reproduce the mass of the $Y(4230)$, the lowest vector tetraquark mass up to now. In the present  work, we extend our previous works   to investigate the three-body strong decays $Y(4230) \to \bar{D}^{*-}D^{*0}\pi^+$, $\bar{D}^{*-}D^0\pi^+$,  $J/\psi\pi^+\pi^- $ and  $ J/\psi K^+K^-$ with the light-cone QCD sum rules by assuming contact four-meson coupling constants. We introduce free parameters to parameterize the higher resonance contributions to acquire  rigorous quark-hadron  duality, and obtain four QCD sum rules for the hadronic coupling constants, then we vary the free parameters to obtain flat Borel platforms therefore extracting the values of the four-meson hadronic coupling constants.  Finally, we obtain the corresponding partial decay widths, which are too small to account for the experimental data. We expect that those decays take place through an intermediate meson to outcome the dilemma, we can search for the intermediate states and precisely measure the branching fractions, which maybe shed light on the nature of the $Y$ states. Furthermore, we expect to search for the vector tetraquark states in the two-body strong decays $Y\to D\bar{D}$, $D^*\bar{D}^*$, $D\bar{D}^*$, $D^*\bar{D}$, as they would take place through the Okubo-Zweig-Iizuka super-allowed fall-apart mechanism.

\section*{Acknowledgements}
This  work is supported by National Natural Science Foundation, Grant Number  12175068.

\end{document}